%% file: ws-ijmpc.tex
\documentclass{ws-ijmpc}
\input{commandsGeneral}

\usepackage{amsmath, amssymb}
\usepackage{tikz}
 
\usepackage{graphicx}
\usepackage{wrapfig}
\usepackage{subfigure}
\usepackage{xcolor}
\usepackage{hyperref}
 
\usepackage{booktabs}

\begin{document}

\markboth{Regina Ammer, Matthias Markl, Vera J\"{u}chter, Carolin K\"{o}rner, Ulrich R\"{u}de}
{Validation Experiments for LBM Simulations of Electron Beam Melting}

\catchline{}{}{}{}{}
%%%%%%%%%%%%%%%%%%%%%%%%%%%%%%%%%%%%%%%%%%%%%%%%%%%%%%%%%%%%%%%%%%%%

\title{VALIDATION EXPERIMENTS FOR LBM SIMULATIONS OF ELECTRON BEAM MELTING}

\author{REGINA AMMER$^{\ast}$ and ULRICH R\"{U}DE$^{\dagger}$ }

\address{Department of Computer Science 10 
System Simulation, University of Erlangen-Nuremberg, Cauerstr. 11, 91058 Erlangen, 
Germany\\$^{\ast}$regina.ammer@fau.de \\$^{\dagger}$ulrich.ruede@fau.de}

\author{MATTHIAS MARKL$^{\star}$, VERA J\"{U}CHTER$^{\dagger}$ and CAROLIN K\"{O}RNER$^{\ddagger}$}

\address{Chair of Metals Science and Technology, University of Erlangen-Nuremberg,\\ Martensstr. 5, 
91058 Erlangen, Germany\\
$^{\star}$matthias.markl@fau.de\\$^{\dagger}$vera.juechter@ww.uni-erlangen.de\\$^{\ddagger}$carolin.koerner@ww.uni-erlangen.de }

\maketitle

\begin{history}
\received{Day Month Year}
\revised{Day Month Year}
\end{history}

\begin{abstract}
\input{abstract.tex}
\end{abstract}

\ccode{PACS Nos.: 11.25.Hf, 123.1K}

\section{Introduction}
\input{introduction.tex}

\section{Numerical methods}
\input{num_methods.tex}

\section{Validation experiments}
\input{validation_ex.tex}
\section{Conclusion}
\input{conclusion.tex}

%%%%%%%%%%%%%%%%%%%%%%%%%%%%%%%%%%%%%%%%%%%%%%%%%%%%%%%%%%%%%%%%%%%%
\end{document}

%% file: commandsGeneral.tex
\newcommand{\walberla}{\textsc{waLBerla}}
\DeclareMathAlphabet{\mathpzc}{OT1}{pzc}{m}{it}

%% file: abstract.tex
This paper validates 3D simulation results of electron beam melting (EBM) processes comparing experimental 
and numerical data. 
The physical setup is presented which is discretized by a three dimensional (3D) thermal lattice Boltzmann method (LBM). 
An experimental process window is used for the validation depending on the line energy injected into the metal 
powder bed and the scan velocity of the electron beam. In the process window the EBM products are classified into
the categories, porous, good and swelling, depending on the quality of the surface. The same parameter sets are used to generate 
a numerical process window. A comparison of numerical and experimental process windows shows a good agreement.
This validates the EBM model and justifies simulations for future improvements of EBM processes. 
In particular numerical simulations can be used to explain future process window scenarios 
and find the best parameter set for a good surface quality and dense products.


%% file: introduction.tex
Electron beam melting (EBM) is an additive manufacturing method used to produce complex metallic structures layer by layer from metal powder~\cite{Heinl2007}. 
EBM opens new opportunities in many industries, ranging from aircraft manufacturers to producers of medical implants.
However, until now the parts cannot be manufactured at sufficient speed to make them economically viable for any but specific very high value applications. 
In order to accelerate the building process and improve the accuracy a better understanding of the beam-powder interaction is necessary. 
This can be gained by 3D simulations of the process.

Based on the work of K\"{o}rner et al.\cite{Koerner2011} we use a 3D thermal LBM for the discretization for the EBM processes. This 
model includes hydrodynamic effects, like melt flow, capillarity and wetting, as well as thermal effects, like absorption of the 
beam energy, melting and solidification. Detailed information for parallel, optimized 3D absorption algorithms is found in~\cite{Markl2013} 
and the description of the modeling aspects, e.g., the generation of the Ti-Al6-V4 powder particles is given in~\cite{Ammer2013}. 
The implementation of the model is embedded in the \walberla{} framework (widely applicable lattice Boltzman solver
from Erlangen) which is a lattice Boltzmann based fluid flow solver with highly parallelized kernels~\cite{walberla2011,kostler2013cse}. 
In~\cite{Ammer2013}  basic validation examples are already shown, like the test of the Stefan problem as benchmark for the simulation of 
solid-liquid phase transition, and modest EBM process examples, e.g., the melting of a spot regarding the interaction of powder particles
and electron beam. In this paper we extend the validation examples focusing on the EBM processes, e.g, hatching scenarios, and compare
the numerical results with experimental data. These numerical results demonstrate the good potential of the thermal LBM approach to 
understand and predict complex processes like the EBM depending on thermodydnamic as well as on fluid dynamic phenomena. 

The remainder of the paper is organized as follows: Section 2 describes the 3D thermal LBM used for the simulation of the 
EBM process. The next section defines line energy, scan velocity, porosity and swelling before showing the experimental
process window. Subsequently, the generation of the numerical process window is described, shown and compared with the 
experimental one. Section 4 concludes the validation experiments and outlines future research topics.

%% file: num_methods.tex
Based on the lattice Boltzmann equation the isothermal lattice Boltzmann method (LBM) is introduced by~\cite{McNamara1988} 
as an improvement of its predecessor, the lattice gas automata (LGA)~\cite{Hasslacher1986}. 
LBM is an alternative, mesoscopic approach for the numerical solution of the Navier-Stokes equations~\cite{Higuera1989} 
and can be parallalized easier than other traditional CFD methods whose computation is based on the conservation of macroscopic quantities. 
LBM has also less computational storage and work requirements than molecular dynamic methods which consider microscopic physics. 

Thermal LB methods are classified into three categories: the multispeed LBM~\cite{Alexander1993,ChenOhashi1994}, 
the double-distribution or multidistribution approach~\cite{Massaioli1993,Shan1997,He1998} and hybrid approach 
where the scalar temperature equation is solved by finite differences or finite volume methods~\cite{Lallemand2003}. 
The multispeed LBM can be seen as an extension of the isothermal LBM where the equilibrium distribution function depends also 
on the temperature and has higher nonlinear velocity terms. This method suffers from numerical instabilities which 
have to be stabilized as it is shown in~\cite{McNamara1995}. An additional disadvantage is that only one Prandtl 
number can be simulated by multispeed LBM which limits the range of applications. The multidistribution LBM overcomes 
these drawbacks using a separate distribution function for the temperature field.  

For the simulation of the EBM process a thermal two-distribution LBM is used which is described followed by 
the treatment of the free surface boundary condition. 
\subsection{3D thermal lattice Boltzmann method}
The model for the EBM process is based on the work of K\"{o}rner et al. in~\cite{Koerner2011} where a 2D thermal LBM is discussed. 
The idea of LBM is solving the Boltzmann equation in the hydrodynamic limit for a particle distribution function (pdf) in the physical momentum space. 
This pdf $f(\textbf{x},\textbf{v},t)$ indicates the probability of finding a 
particle with the macroscopic velocity $\textbf{v}(\textbf{x},t)$ at position $\textbf{x}$ and time $t$. For the thermal LBM a second pdf set 
$h(\textbf{x},\textbf{v},t)$ is used for the numerical solution of the temperature field.
The collision operator is approximated by the BGK-collision operator~\cite{BGK1954,ChenOhashi1994},
\begin{equation}
 \Omega_{f}=- \frac{1}{ \tau_{f} }\left( f-f^{eq} \right), \quad \Omega_{h} = - \frac{1}{ \tau_{h} }\left( h-h^{eq} \right),
\end{equation}
where $f^{eq}$ and $h^{eq}$ denote the Maxwell equilibrium distribution~\cite{HeLuo1997}. $\tau_{f}$ is the relaxation time 
related to the viscosity $\nu$ of the fluid and given by $\nu = c_{s}^{2}\Delta t(\tau_{f} -0.5)$ with the lattice speed of sound 
$c_{s}^{2}= \frac{\Delta x^{2}}{3\Delta t^{2}}$. $\tau_{h}$ belongs to the second pdf set $h_{i}$ and there exists the relation to 
the thermal diffusivity $k = c_{s}^{2}\Delta t (\tau_{h} - 0.5)$. 

The D3Q19 stencil is used to discretize the microscopic space, i.e., a finite set of 19 discrete velocities $\textbf{e}_{i}$ 
and lattice weights $\omega_{i}$ are given~\cite{HeLuo1997}. The Maxwell equilibrium distribution is discretized by a Taylor series 
expansion~\cite{HeLuo1997_} and is given by,
\begin{equation}
 f_{i}^{eq}(\textbf{x},t) = \omega_{i} \rho \left[ 1 + \frac{ (\textbf{e}_{i} \cdot \textbf{u}) }{ c_{s}^{2}} + \frac{ ( \textbf{e}_{i} \cdot \textbf{u} )^{2} }{2 c_{s}^{4}} - \frac{ \textbf{u}^{2} }{ 2 c_{s}^{2} }  \right],\,  h_{i}^{eq}( \textbf{x},t ) = \omega_{i} E \left[ 1 + \frac{ (\textbf{e}_{i} \cdot \textbf{u} ) }{ c_{s}^{2} } \right]
\end{equation}
where $\textbf{u}$ denotes the velocity of the liquid. It is sufficient to use a linearized equilibrium distribution function $h_{i}^{eq}$ for the temperature field. 
The macroscopic quantities density $\rho$, momentum $\rho \textbf{u}$ and energy density $E$ are computed in the following way,
\begin{equation}
 \rho = \sum \limits_{i} f_{i} ,\quad \rho \textbf{u} = \sum \limits_{i} \textbf{e}_{i}f_{i}, \quad E = \sum \limits_{i} h_{i}.
\end{equation}
Both pdf sets are one-way coupled by the velocity of the fluid $\textbf{u}$ which is used for the computation of $h_{i}^{eq}$. 
The discretization of the EBM process by a 3D coupled-thermal LBGK~\cite{Ammer2013} can be written in a stream-collide algorithm,
\begin{align}\operatorname{streaming}
 \label{eq:thermal_streaming}
 \begin{cases}
 f_{i}^{'}(\textbf{x}+\textbf{e}_{i}\Delta t, t+ \Delta t)&= f_{i}(\textbf{x},t) \\
 h_{i}^{'}(\textbf{x}+\textbf{e}_{i}\Delta t, t+ \Delta t)&= h_{i}(\textbf{x},t),
 \end{cases} 
\end{align}
\begin{align} \operatorname{collide}
 \label{eq:thermal_collide}
  \begin{cases}
   f_{i}(\textbf{x},t + \Delta t)  &= f_{i}^{'} (\textbf{x},t+\Delta t) + \frac{1}{\tau_{f}} ( f_{i}^{eq}(\rho,\textbf{u}) - f_{i}(\textbf{x},t + \Delta t) ) + F_{i} \\
   h_{i}(\textbf{x},t + \Delta t)  &= h_{i}^{'} (\textbf{x},t+\Delta t) + \frac{1}{\tau_{h}} ( h_{i}^{eq}(E,\textbf{u}) - h_{i}(\textbf{x},t + \Delta t) ) + \Phi_{i},
  \end{cases}
\end{align}
where $F_{i}$ denotes an external force term, e.g., gravity, and $\Phi_{i}$ is the energy injected into the system by the electron 
beam,
\begin{equation}
 F_{i} = \omega_{i}\rho \left[ \frac{ (\textbf{e}_{i}-\textbf{u}) }{c_{s}^{2}}  + \frac{ (\textbf{e}_{i} \cdot \textbf{u})\textbf{e}_{i}  }{c_{s}^{4}} \right]\cdot \textbf{g}, \quad \Phi_{i}(\textbf{x},t) = \omega_{i}E_{b}(\textbf{x},t),
\end{equation}
where $E_{b}$ denotes the energy of the electron beam. Detailed information how the beam energy and absorption are computed is given in~\cite{Markl2013,Ammer2013}. Next, 
the free surface boundary condition is outlined. 

\subsection{Free surface lattice Boltzmann method }
K\"{o}rner et al. introduce in~\cite{Koerner2005} a volume-of-fluid based free surface lattice Boltzmann approach. This method neglects the gas phase and 
works completely on the liquid phase assuming that the liquid phase covers the behavior of the flow completely. 
In order to ensure mass and momentum conservation a boundary condition is imposed at the interface
which separates liquid and gas phase. 
More information regarding the free surface boundary condition used for this EBM model can be found in~\cite{Ammer2013}. 

% K\"{o}rner et al. introduce the free surface boundary condition in \cite{Koerner2005}. This Volume-of-Fluid based approach neglects the gas phase and works completely on the 
% liquid phase assuming that the liquid phase covers the behavior of the flow completely. The two phases are separated by an interface layer where the cells are partially filled defined by 
% the fill level. The free surface boundary condition postulates that the force evoked by the liquid has to be the same as the gas force. Therefore,  pdfs - coming from the gas phase - 
% have to be reconstructed. This reconstruction process for the missing pdf of the gas phase fulfills the force boundary condition based on a momentum exchange method. Gas forces are known by
% the gas pressure as well as the velocity computed at the interface. This free surface boundary condition assures that the macroscopic quantities are conserved and can be computed although the 
% gas phase is not computed explicitly. More information regarding the free surface boundary condition used for the EBM model can be found in~\cite{Ammer2013}. 
%
%
%\subsection{Implementation}

%% file: validation_ex.tex
The following validation experiments do not only cover particular aspects as the solid-liquid phase transition or only concentrate 
on qualitative accordance but consider also the complete EBM process in a quantitative way. 
\subsection{Definitions and experimental setup}
We use the following experimental setup for the validation experiments. 
\begin{wrapfigure}{r}{5cm}
\centering
\vspace{-13pt}
\includegraphics[width=0.4\textwidth]{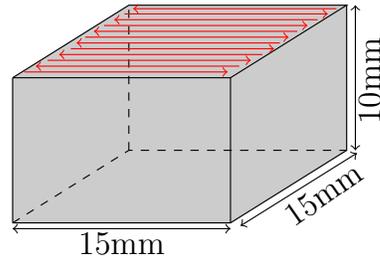}
% \begin{tikzpicture}[scale=0.35]
%   % fill grey
%   \filldraw[black!20] (0,0)--(7.5,0)--(11.5,2.5)--(4,2.5)--(0,0);
%   \filldraw[black!20] (0,5)--(7.5,5)--(11.5,7.5)--(4,7.5)--(0,5);
%   \filldraw[black!20] (0,0)--(7.5,0)--(7.5,5)--(0,5)--(0,0);
%   \filldraw[black!20] (7.5,0)--(11.5,2.5)--(11.5,7.5)--(7.5,5)--(7.5,0);
%   % draw grid
%   \draw (0,0)--(7.5,0)--(11.5,2.5);
%   \draw[dashed] (0,0)--(4,2.5)--(11.5,2.5);
%   \draw (0,5)--(7.5,5)--(11.5,7.5)--(4,7.5)--(0,5);
%   \draw (0,0)--(0,5);
%   \draw (7.5,0)--(7.5,5);
%   \draw (11.5,2.5)--(11.5,7.5);
%   \draw[dashed] (4,2.5)--(4,7.5);
%   % draw beam
%   \draw[red,thin,->](0.4,5.2)--(7.4,5.2);
%   \draw[red,thin,<-](0.8,5.4)--(7.8,5.4);
%   \draw[red,thin,->](1.2,5.6)--(8.2,5.6);
%   \draw[red,thin,<-](1.6,5.8)--(8.6,5.8);
%   \draw[red,thin,->](2.0,6.0)--(9.0,6.0);
%   \draw[red,thin,<-](2.2,6.2)--(9.2,6.2);
%   \draw[red,thin,->](2.5,6.4)--(9.5,6.4);
%   \draw[red,thin,<-](2.8,6.6)--(9.8,6.6);
%   \draw[red,thin,->](3.15,6.8)--(10.15,6.8);
%   \draw[red,thin,<-](3.5,7.0)--(10.5,7.0);
%   \draw[red,thin,->](3.85,7.2)--(10.85,7.2);
%   \draw[red,thin,<-](4.15,7.38)--(11.15,7.38);
%   % nodes, declaration
%   \draw[thin,<->] (0,-0.2)--(7.5,-0.2);
%   \node at (3.75,-0.70) {15mm};
%   \draw[thin,<->] (7.85,-0.1)--(11.85,2.4);
%   \node[rotate=35] at (10.7,1.0) {15mm};
%   \draw[thin,<->] (11.8,2.5)--(11.8,7.5);
%   \node[rotate=90] at (12.3,4.9) {10mm};
% \end{tikzpicture}
%\vspace{-13pt}
\caption{Experimental setup.}
%\vspace{-13pt}
\label{fig:exp_setup}
\end{wrapfigure}
A cuboid of size (15x15x10)\,mm$^{3}$ is generated by hatching the Ti-Al6-V4 powder particles layer by layer (compare the red arrows in Fig.~\ref{fig:exp_setup}).
The hatching differs by line energy and scan velocity of the electron beam. The line energy is defined by,
\begin{equation}
 E_{L} = \frac{U\cdot I}{v_{\text{scan}}} = \frac{P_{\text{beam}}}{v_{\text{scan}}},
\label{eq:line_energy}
\end{equation}
where $U$ denotes the acceleration voltage in [V], $I$ the current in [A] and $v_{\text{scan}}$ the scan velocity in [$\frac{\text{m}}{\text{s}}$] of the electron beam. 
The parameter set $(E_{L},v_{\text{scan}})$ defines the electron beam. 
The quality of EBM products is classified into three categories, namely porous, good and swelling. Good samples have a smooth surface and 
a relative density higher than 99.5\%. If the temperature during the process is too high swelling can occur and the dimensional accuracy cannot be guaranteed. 
\begin{figure}[htbp!]
\centering
\includegraphics{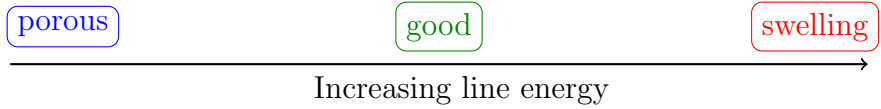}
%\vspace{-13pt}
%\begin{tikzpicture}
% \node[blue,rectangle,rounded corners,draw] at (0,0) {porous};
% \node[green!50!black,rectangle,rounded corners,draw] at (5,0) {good};
% \node[red,rectangle,rounded corners,draw]  at (10,0) {swelling};
% \draw[thick,->] (-0.7,-0.5)--(10.7,-0.5);
% \node at (5.3,-0.85) {Increasing line energy $E_{\text{L}}$};
%\end{tikzpicture}
%\vspace{-8pt}
\caption{Categories of samples.}
\label{fig:categories}
%\vspace{-13pt}
\end{figure}
On the other side, if the line energy is too low and the relative density is smaller than
99.5\%, then the sample is porous (cf. Fig.~\ref{fig:categories}). 
Next, a experimental process window is shown in Fig.~\ref{fig:pw_real} which is compared with numerical simulation results. 

\subsection{Experimental data}
\begin{figure}[htbp!]
 \centering
 \includegraphics[width=0.7\textwidth]{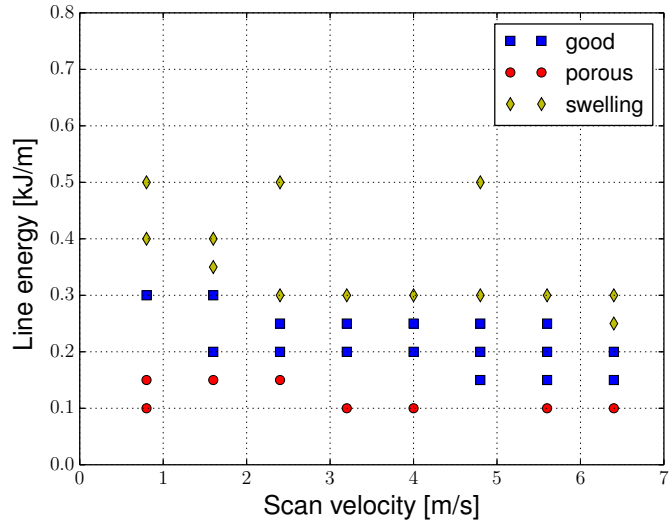}
 \caption{Process window of experimental data. Built temperature 1000\,K, line offset 100\,$\mu$m, layer thickness 50\,$\mu$m, focused electron beam.}
 \label{fig:pw_real}
\end{figure}

Fig.~\ref{fig:pw_real} shows the quality of samples with different line energies in [$\frac{\text{kJ}}{\text{m}}$] and scan velocities in [$\frac{\text{m}}{\text{s}}$] of the electron beam. 
The red circles stand for a porous parameter set, the blue squares for an optimal, good parameter set and the yellow rhombus for a sample where 
swelling occurs because of too high temperatures at the surface. It can be observed that the higher the scan velocity the lower the embedded line energy has to be to get an optimal sample
result. 

\subsection{Numerical results}
We validate our 3D numerical EBM model against these experimental data of Fig.~\ref{fig:pw_real}. 
Because of the high computational cost of the 3D simulations we only model the hatching of one powder 
layer instead of the multiple layers.
\begin{wrapfigure}{r}[0cm]{6.4cm}
%\vspace{-18pt}
 \centering 
 \includegraphics[width=0.5\textwidth]{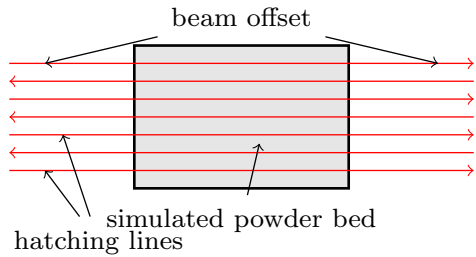}
%  \begin{tikzpicture}[scale=0.38]
%   \filldraw[black!10] (0.5,0.5)--(6.5,0.5)--(6.5,4.5)--(0.5,4.5)--(0.5,0.5);
%   \draw[thick] (0.5,0.5)--(6.5,0.5)--(6.5,4.5)--(0.5,4.5)--(0.5,0.5);
%   \draw[red,->] (-3,1.0)--(10,1.0);
%   \draw[red,<-] (-3,1.5)--(10,1.5);
%   \draw[red,->] (-3,2.0)--(10,2.0);
%   \draw[red,<-] (-3,2.5)--(10,2.5);
%   \draw[red,->] (-3,3.0)--(10,3.0);
%   \draw[red,<-] (-3,3.5)--(10,3.5);
%   \draw[red,->] (-3,4.0)--(10,4.0);
%   \node(powderbed) at (3.5,-0.4) {\scriptsize{simulated powder bed}};
%   \draw[->] (powderbed)--(4,1.75);
%   \node(hatchinglines) at (-0.5,-1) {\scriptsize{hatching lines}};
%   \draw[->] (hatchinglines)--(-2,1);
%   \draw[->] (hatchinglines)--(-1.5,2);
%   \node(beamoffset) at (3.5,5.3) {\scriptsize{beam offset}};
%   \draw[->] (beamoffset)--(-2,4);
%   \draw[->] (beamoffset)--(9,4);
%  \end{tikzpicture}
 \caption{Sketch of simulation scenario.}\label{fig:SketchSimScenario}
 \vspace{-10pt}
\end{wrapfigure} 
One simplified scenario for the exemplary parameter set of $(3.2\,\frac{\text{m}}{\text{s}},0.2\,\frac{\text{J}}{\text{m}})$ does already require 8 compute 
nodes of \textsc{LiMa}\footnote{http://www.rrze.uni-erlangen.de/dienste/arbeiten-rechnen/hpc/systeme/lima-cluster.shtml} for 4 wall clock hours.  
\textsc{LiMa} is the compute cluster where all 3D simulations are done. However, this simplification is in good agreement with the situation
expected for the full setup as it will be shown in this Section. 
We also minimize the simulated powder particle layer, i.e., we focus only a rectangular powder layer and have seven hatching 
lines (cf. Fig.~\ref{fig:SketchSimScenario}). Thus, a beam offset per layer is defined where the electron beam is on but outside 
the simulated powder particle bed. 

\begin{figure}[htpb!]
\centering
 \includegraphics[width=0.7\textwidth]{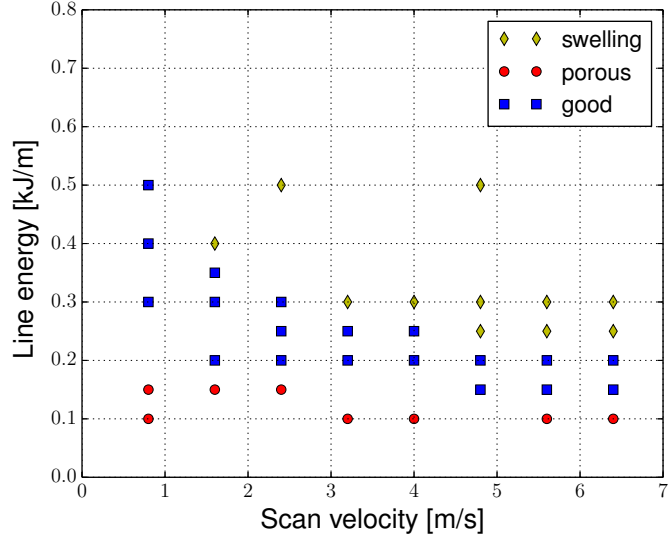}
 %\vspace{-8pt}
 \caption{Process window of numerical data. $\Delta x=5\cdot10^{-6}$\,m, $\Delta t = 1.75\cdot10^{-7}$\,s, simulated powder domain (1.44x0.64x0.24)$\cdot10^{-3}\,\text{m}^{3}$, beam offset =13.56$\cdot10^{-3}$\,m.}
 %\vspace{-13pt}
 \label{fig:pw_num}
\end{figure}

In Fig.~\ref{fig:pw_num} the quality of the numerical experiments with different line energy and scan velocity is shown. 
In order to determine if the numerical sample is porous the relative density of the sample is measured and if it is less than 99.5\% the 
sample is porous. The definition of numerical swelling is more difficult. Our model does not include evaporation and thus no evaporation pressure. 
Hence, numerical swelling effects cannot occur. We have to find another way how we can determine regions numerically where swelling exists experimentally. 
Therefore, we use the numerical computed temperature which is higher than in real EB melting processes because the cooling effect during the process of evaporation is missing.  
Furthermore, numerical artefacts cause outliers in the temperature field. 
In order to overcome these numerical influences we use an averaging process for the temperature values. If these averaged values are still higher than 7500\,K 
we assume that swellings occur in the experimental setup.

%
%\vspace{-13pt}
In comparison with the experimental data in Fig.~\ref{fig:pw_real} it is observed that all porous samples 
are consistent in the simulation results (cf. Fig.~\ref{fig:pw_num}). Furthermore, the numerical and experimental results for scan velocities 
of 3.2\,$\frac{\text{m}}{\text{s}}$, 4.0\,$\frac{\text{m}}{\text{s}}$ and 6.4\,$\frac{\text{m}}{\text{s}}$ agree with each other.
Small differences between numerical and experimental data are only found for 0.8\,$\frac{\text{m}}{\text{s}}$, 1.6\,$\frac{\text{m}}{\text{s}}$, 2.4\,$\frac{\text{m}}{\text{s}}$, 
4.8\,$\frac{\text{m}}{\text{s}}$ and 5.6\,$\frac{\text{m}}{\text{s}}$ for higher line energies; for $v_{\text{scan}}=$~0.8\,$\frac{\text{m}}{\text{s}}$ in experimental data swelling
already occur for 0.4 and 0.5\,$\frac{\text{kJ}}{\text{m}}$ but in numerical data no swelling is measured. The same effect is seen for scan velocities 1.6 and 2.4\,$\frac{\text{m}}{\text{s}}$ 
where swelling occurs for smaller line energies experimentally than numerically. % 
%for line energies of 0.35 and 0.4\,$\frac{\text{kJ}}{\text{m}}$ and 0.3 and 0.5\,$\frac{\text{kJ}}{\text{m}}$, respectively, in experiments and in numerical 
%simulations swelling only exists for line energies of 0.4\,$\frac{\text{kJ}}{\text{m}}$ and 0.5\,$\frac{\text{kJ}}{m}$. 
For higher scan velocities of 4.8 and 5.6\,$\frac{\text{m}}{\text{s}}$ 
the reverse effect is observed, i.e., more swelling occur numerically than experimentally. For $v_{\text{scan}}=4.8\,\frac{\text{m}}{\text{s}}$ swelling arises numerically already for a line 
energy of 0.25\,$\frac{\text{kJ}}{\text{m}}$ while the experimental setup is still good. The same is seen for $v_{\text{scan}}=5.6\,\frac{\text{m}}{\text{s}}$; line energies higher than
0.25\,$\frac{\text{kJ}}{\text{m}}$ lead to swelling in numerical simulations, but in experimental data only energies equal 0.3\,$\frac{\text{kJ}}{\text{m}}$ arise swelling. 

Both effects can be explained by the focus of the electron beam gun. The focus is more precise for smaller scan velocities imposed by a smaller electron beam power $P_{\text{beam}}$, e.g, 
$v_{\text{scan}}=1.6\,\frac{\text{m}}{\text{s}}$ or $v_{\text{scan}}=2.4\,\frac{\text{m}}{\text{s}}$ lead to a precise, small focus and thus, energy is brought onto a smaller area. 
For these velocities swelling occurs for smaller line energies experimentally than numerically because the EB gun focus is constant for different beam power and scan velocities, respectively. 
For higher scan velocities (larger $P_{\text{beam}}$) 
like 4.8\,$\frac{\text{m}}{\text{s}}$ the focus of the gun spreads and, subsequently, the same amount of energy is brought into a larger area of powder particles and the maximum 
temperature is smaller. As a consequence in experimental data less swelling occurs for higher scan velocities (larger $P_{\text{beam}}$) than in numerical data. 

\begin{figure}[htpb]
\vspace{-15pt}
\centering
 \subfigure[t =  0.20$\cdot10^{-3}$\,s \label{fig:hatching_1}]{\includegraphics[width = 0.3\textwidth]{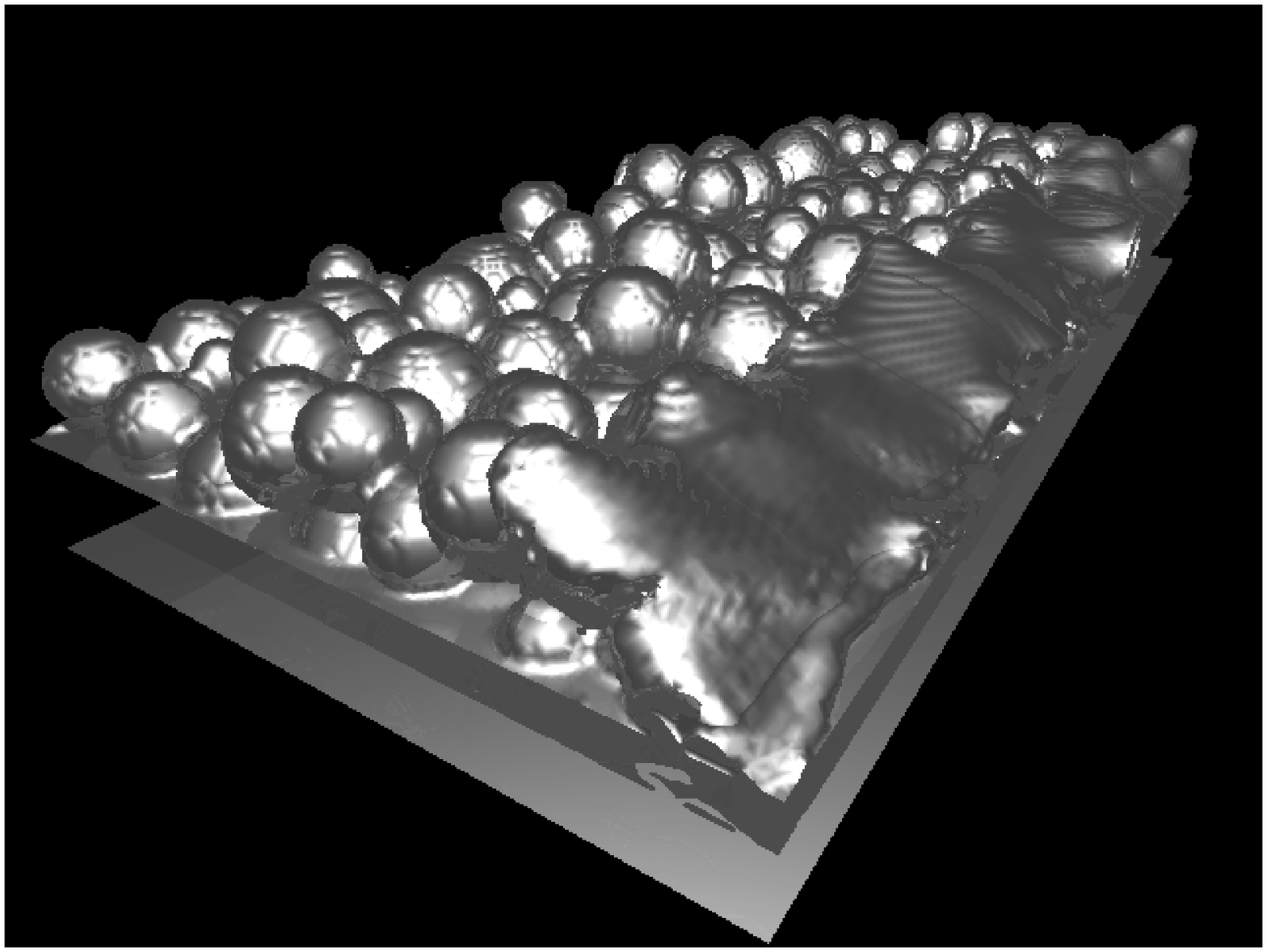}}  \hspace{2ex}
 \subfigure[t =  4.27$\cdot10^{-3}$\,s \label{fig:hatching_2}]{\includegraphics[width = 0.3\textwidth]{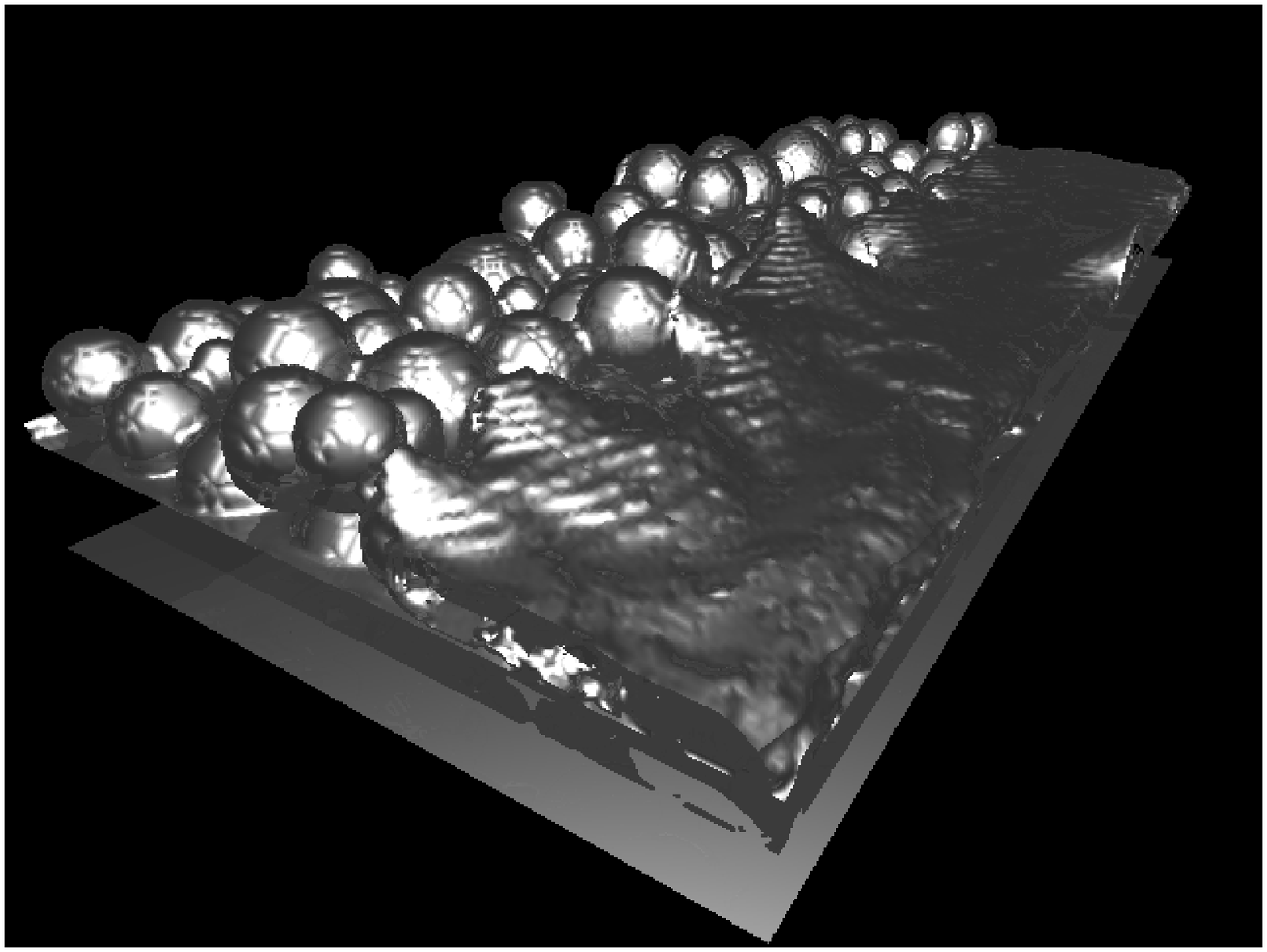}}  \hspace{2ex}
 \subfigure[t = 12.91$\cdot10^{-3}$\,s \label{fig:hatching_3}]{\includegraphics[width = 0.3\textwidth]{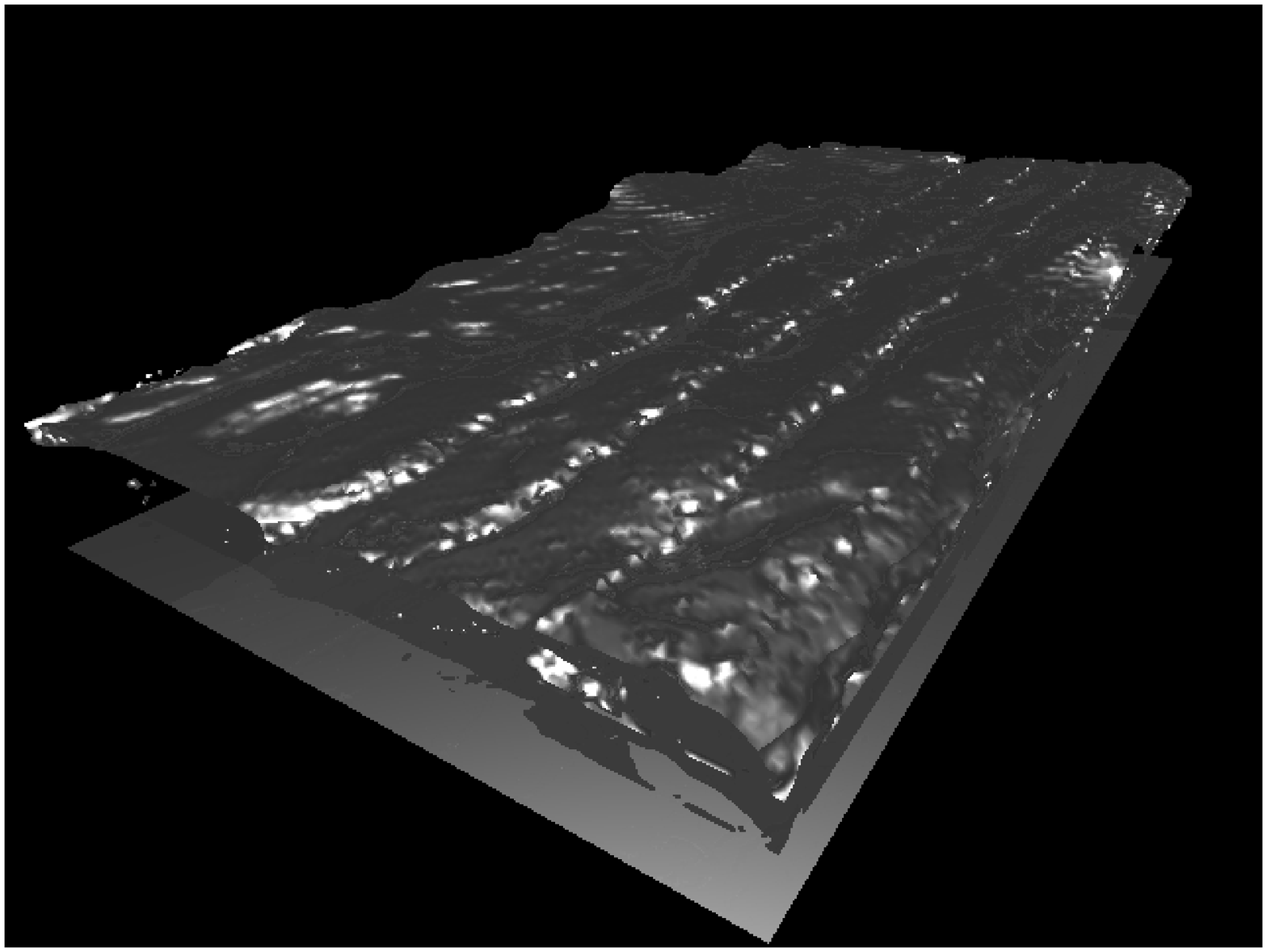}} \hspace{2ex}
 \vspace{-13pt}
\caption{Hatching of one layer, (0.15$\,\frac{\text{kJ}}{m}$, 6.4\,$\frac{\text{m}}{\text{s}}$).}
\label{fig:Hatching}
\vspace{-15pt}
\end{figure}
Fig.~\ref{fig:Hatching} shows hatching of one line with parameter set (0.15$\frac{\text{kJ}}{\text{m}}$, 6.4\,$\frac{\text{m}}{\text{s}}$) for three 
different time steps. The inverse Gaussian distributed powder particles~\cite{Ammer2013} are demonstrated the  and the free surface is visualized by an isosurface. 
In Fig.~\ref{fig:hatching_1} the electron beam melted one line and powder particles in this first line are not yet completely melted. 
The next Fig.~\ref{fig:hatching_2} shows the second melted line by a coming-back 
electron beam. Here, the particles of the first line are now melted completely. In the third Fig.~\ref{fig:hatching_3} the electron beam has melted 
seven hatching lines and the whole powder layer is melted. 

%% file: conclusion.tex
% conclusion
This paper describes the 3D thermal LB method for the numerical simulation of EBM processes. 
Experimental data of hatching a cuboid with different line energies and scan velocities are shown in a process window. The different 
samples are classified into porous, good and swelling, depending on the parameter set (line energy, scan velocity). For the validation all parameter sets are tested 
numerically and a numerical process window is generated by 3D simulations of hatching one powder layer. Numerical and experimental 
process window are highly concordant especially in the region of small line energies because parameter sets which are porous in experimental 
process window are that also in the numerical one. Only in the region between good and swelling differences exist which are 
reducible to the focus of the EB gun. The precision of the focus depends on the beam power but in our simulations the focus is constant. 

In order to accelerate the building process of EBM manufacturing it is important to extend the process window, i.e., to raise up the 
question how the process window will look like for a higher beam power and therefore higher scan velocities. After the validation experiments in this paper which 
justify the correct behavior of our framework and its beneficing to simulate EBM processes we will try to find better parameter 
sets regarding quality of EBM products. 

\section*{Acknowledgments}
Our work is supported by the European Union Seventh Framework Program -- Research for SME's with full title 
``High Productivity Electron Beam Melting Additive Manufacturing Development for the Part Production Systems Market'' and 
grant agreement number 286695. We also gratefully thank the German Research Foundation (DFG) for funding the experimental work within the  Collaborative Research Centre 814, project B2. 
%The authors of this paper would like to thank all developers of \walberla{} and \pe{} and all 
%material scientists responsible for the real experiments. 

%% file: ws-ijmpc.bbl
\begin{thebibliography}{00}
% Selective Manufacturing with EBM - application!!!
\bibitem{Heinl2007}
P. Heinl et al., % Andreas Rottmair, Carolin K\"{o}rner and Robert F. Singer,
{\it Adv. Eng. Mater.} {\bf 9} 360 (2007).

\bibitem{Koerner2011}
C. K\"{o}rner et al.,
%{\it Journal of Materials Processing Technology} {\bf 211} 978 (2011).
{\it J. Mater. Process. Tech.} {\bf 211} 978 (2011).

\bibitem{Markl2013}
M. Markl et al.,
%{\it Procedia Computer Science} {\bf 18} 2127 (2013). 
{\it Procedia Comput. Sci.} {\bf 18} 2127 (2013). 

\bibitem{Ammer2013}
R. Ammer et al,
%{\it Computers and Mathematics with Applications} {\bf } p.~ .(2013) 
{\it Comput. Math. Appl.} {\bf 67} 318 (2014)


%%%%%%%%%%%%%%%%%%%%%%%%%%%%%%%%%%%%%%%%%%%%%%%%%%%%%%%%%
% HPC + LSS
\bibitem{walberla2011}
C. Feichtinger et al., %S. Donath, H. K\"{o}stler, J. G\"{o}tz, U. R\"{u}de,
%WaLBerla: HPC software design for computational engineering simulations,
%{\it Journal of Computational Science} {\bf 2} 105 (2011). % p.~105-112
{\it J. Comput. Sci.} {\bf 2} 105 (2011). % p.~105-112

\bibitem{kostler2013cse}
H. K\"{o}stler et al., %and U. R\"{u}de,
%The CSE Software Challenge—Covering the Complete Stack.
{\it IT-Information Technology} {\bf 55} 91 (2013). % p.~91-96. 

\bibitem{McNamara1988}
G.R. McNamara et al.,
%Use of the Boltzmann equation to simulate lattice-gas automata.
{\it Phys. Rev. Lett.} {\bf 61} 2332 (1988). % p.~2332-2335. 

\bibitem{Hasslacher1986}
U. Frisch et al., % B. Hasslacher and Y. Pomeau,
%Lattice - Gas automata for the Navier-Stokes equation.
{\it Phys. Rev. Lett.} {\bf 56 } 1505 (1986). % p.~1505-1508.

\bibitem{Higuera1989}
F. Higuera et al., %and J. Jimenez,
%Boltzmann approach to lattice gas simulations.
{\it Europhys. Lett.} {\bf 9} 663 (1989) %p.~663-668.

\bibitem{Alexander1993}
F.J. Alexander et al., % S.Chen and J.D. Sterling,
% Lattice Boltzmann thermodhydrodynamics.
{\it Phys. Rev. E} {\bf 47} R2249 (1993) 

\bibitem{ChenOhashi1994}
Y. Chen et al., % H. Ohashi and M. Akiyama,
% Thermal lattice Bhatnagar-Gross-Krook model without nonlinear deviations in macrodynamic equations
{\it Phys. Rev. E} {\bf 50} 2776 (1994)

\bibitem{Massaioli1993}
F. Massaioli et al., % R. Benzi and S. Succi,
%Exponential Tails in Two-dimensional Rayleigh-B´enard Convection.
{\it Europhys. Lett.} {\bf 21} 305 (1993)

\bibitem{Shan1997}
X. Shan, 
% Simulation of Rayleigh-Benard convection using a lattice Boltzmann method
{\it Phys. Rev. E} {\bf 55} 2780 (1997)

\bibitem{He1998}
X. He et al., %S. Chen and G. Doolen,
%A novel thermal model for the lattice Boltzmann method in incompressible limit.
{\it J. Comput. Phys.} {\bf 146} 282 (1998). % p.~282 - 200. 

\bibitem{Lallemand2003}
P. Lallemand et al., % L. Luo,
% Hybrid finite-difference thermal lattice Boltzmann equation
{\it Int. J. Mod. Phys B} {\bf 17} 41 (2003)

\bibitem{McNamara1995}
G.R. McNamara et al., %and B. Alder,
%Stabilization of thermal lattice Boltzmann methods.
{\it J. Stat. Phys.} {\bf 81} 395 (1995). % p.~395-408. 

\bibitem{BGK1954}
P. Bhatnagar et\,al.,
{\it Phys. Rev.} {\bf 94} 511 (1954).

\bibitem{ChenDoolen1998}
S. Chen et al., %and G. Doolen,
%Lattice Boltzmann Method for Fluid Flows.
{\it Annu. Rev. Fluid Mech.} {\bf 30} 329 (1998). % p.~329-364.

\bibitem{HeLuo1997}
X. He et al., % L. Luo,
% Theory of the lattice Boltzmann method: From the Boltzmann equation to the lattice Boltzmann equation
{\it Phys. Rev. E} {\bf 56} 6811 (1997)

\bibitem{HeLuo1997_}
X. He et al., % L. Luo,
% A priori derivation of the lattice Boltzmann equation.
{\it Phys. Rev. E} {\bf 55} R6333 (1997)

\bibitem{Koerner2005}
C. K\"{o}rner et al., %M. Thies, T.Hofmann, N. Th\"{u}rey and Ulrich R\"{u}de,
%Lattice Boltzmann Model for Free Surface Flow for Modeling Foaming.
{\it J. Stat. Physics} {\bf 121} 179 (2005). % p.~179-196.


%%%%%%%%%%%%%%%%%%%%%%%%%%%%%%%%%%%%%%%%%%%%%%%%%%%%%%%%%%%%%%%%%%%%%%%%%%%%%%%%%%%%%%%%%%%%%%%%%%%%%%%%%%%%%%%%%%%%%%%%
%%%%%%%%%%%%%%%%%%%%%%%%%%%%%%%%%%%%%%%%%%%%%%%%%%%%%%%%
% LBM Basics







% 

%%%%%%%%%%%%%%%%%%%%%%%%%%%%%%%%%%%%%%%%%%%%%%%%%%%%%%%%%
% free surface









%%%%%%%%%%%%%%%%%%%%%%%%%%%%%%%%%%%%%%%%%%%%%%%%%%%%%%%%%
%%%%%%%%%%%%%%%%%%%%%%%%%%%%%%%%%%%%%%%%%%%%%%%%%%%%%%%%%
%%%%%%%%%%%%%%%%%%%%%%%%%%%%%%%%%%%%%%%%%%%%%%%%%%%%%%%%%
%\bibitem{}
%{\it } {\bf } () p.~

\end{thebibliography}
